\documentclass[12pt]{iopart}

\usepackage[pdftex]{graphicx}
\usepackage{braket}

\newcommand{\un}[1]{\ensuremath{\,\mathrm{#1}}}
\newcommand{\fig}[1]{figure~\ref{fig:#1}}
\newcommand{\eq}[1]{(\ref{eq:#1})}
\newcommand{\lr}[1]{\ensuremath{\left( #1 \right)}}
\renewcommand{\vec}[1]{\ensuremath{\bi{#1}}}

\renewcommand{\Im}[1]{\ensuremath{\mathrm{Im} \left(#1\right)}}
\newcommand{\I}{\mathrm{i}}

\newcommand{\Gm}{\Gamma}

\newcommand{\Dl}{\Delta}

\newcommand{\om}{\omega}

\begin{document}
\title[From coherent electron focusing to edge channel transport]{Magnetotransport along a boundary: From
  coherent electron focusing to edge channel transport}
\author{T Stegmann, D E Wolf, A Lorke} 
\address{Department of Physics, University of Duisburg-Essen and CENIDE, 47048 Duisburg, Germany} 
\ead{\mailto{thomas.stegmann@uni-due.de}}

\begin{abstract}
  We study theoretically how electrons, coherently injected at one point on the boundary of a two-dimensional
  electron system, are focused by a perpendicular magnetic field $B$ onto another point on the boundary. Using
  the non-equilibrium Green's function approach, we calculate the generalized 4-point Hall resistance $R_{xy}$
  as a function of $B$. In weak fields, $R_{xy}$ shows the characteristic equidistant peaks observed in the
  experiment and explained by classical cyclotron motion along the boundary.  In strong fields, $R_{xy}$ shows
  a single extended plateau reflecting the quantum Hall effect. In intermediate fields, we find superimposed
  upon the lower Hall plateaus anomalous oscillations, which are neither periodic in $1/B$ (quantum Hall
  effect) nor in $B$ (classical cyclotron motion).  The oscillations are explained by the interference between
  the occupied edge channels, which causes beatings in $R_{xy}$. In the case of two occupied edge channels,
  these beatings constitute a new commensurability between the magnetic flux enclosed within the edge channels
  and the flux quantum. Introducing decoherence and a partially specular boundary shows that this new effect
  is quite robust.
\end{abstract}

\submitto{New Journal of Physics \textbf{15} (2013) 113047}
\maketitle

\section{Introduction} \label{sec:Introduction} 

When electrons are injected coherently at one point on the boundary of a two-dimensional electron gas (2DEG),
they can be focused by a perpendicular magnetic field $B$ onto another point of that boundary \cite{Tsoi1999}.
In the classical regime, resonances are observed when a multiple of the cyclotron diameter equals the distance
between the injecting and collecting point contacts. For large Fermi wavelength and long phase coherence
length additional interference effects are observed. This regime of \textit{coherent electron focusing} has
been studied for the first time by van Houten et al. \cite{Houten1989}. Recently, the effects of disorder
\cite{Maryenko2012} and spin-orbit interaction \cite{Usaj2004, Rokhinson2004, Dedigama2006, Reynoso2008,
  Kormanyos2010} were investigated and focusing experiments in graphene were performed
\cite{Taychatanapat2013}. It was also discussed to study by coherent electron focusing the structure of
graphene edges \cite{Rakyta2010} as well as Andreev reflections in normal-superconductor systems
\cite{Polinak2006, Rakyta2007, Haugen2011}. Moreover, a 2DEG in a strong magnetic field shows the quantum Hall
effect, which is explained by the transport through \textit{edge channels} straight along the boundary of the
system \cite{Datta1997}.

\begin{figure}[t]
  \centering
  \includegraphics[scale=1.26]{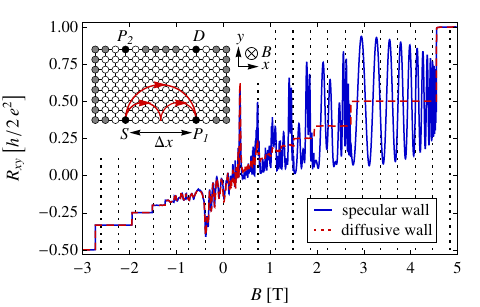}
  \caption{The Hall resistance $R_{xy}=\frac{\mu_{P_1}-\mu_{P_2}}{eI_{SD}}$ as a function of a magnetic field
    $B$ for the system sketched in the inset. In weak fields, $R_{xy}$ shows equidistant focusing peaks
    indicated by dashed vertical lines, when a multiple of the cyclotron diameter equals $\Dl x$. In
    intermediate fields, anomalous oscillations appear, which are neither periodic in $1/B$ nor in $B$.  A
    single Hall plateau is found in large fields, whereas lower Hall plateaus can only by seen when specular
    reflections are suppressed by an absorbing diffusive wall between $S$ and $P_1$.}
  \label{fig:1}
\end{figure}

Although the coherent electron focusing and the quantum Hall effect have been studied extensively in the last
two decades, to our knowledge the two regimes have always been separated. Here, we intermix the two regimes by
suitable system parameters and study theoretically the properties of the focusing experiment emerging at the
transition from the classical cyclotron motion to the quantum Hall edge channel transport. Using the
non-equilibrium Green's function (NEGF) approach \cite{Datta1997, Datta2005, Datta2012}, we calculate the
generalized 4-point Hall resistance $R_{xy}$ as a function of a perpendicular magnetic field $B$. In weak
magnetic fields, the focusing spectrum shows equidistant peaks (see \fig{1}), which can be explained by
classical cyclotron orbits. In strong magnetic fields, the typical fingerprint of the quantum Hall effect can
be observed, i.e. an extended Hall plateau with $R_{xy}=h/2e^2$. Additionally and somewhat unexpectedly, in
intermediate fields, instead of lower Hall plateaus we find oscillations, which are neither periodic in $1/B$
(quantum Hall effect) nor periodic in $B$ (classical cyclotron motion). These oscillations can be explained by
the interference of the occupied edge channels causing beatings in $R_{xy}$.

\section{System} \label{sec:System}

We consider a GaAs-AlGaAs heterojunction, where a 2DEG is formed at the interface of the two
semiconductors. The 2DEG is described by the Hamiltonian
\begin{equation}
  \label{eq:1}
  H_{\mathrm{2DEG}}= \frac{\lr{\vec{p}-e\vec{A}}^2}{2m},
\end{equation}
where $m=0.07m_e$ is the effective mass of the electrons, and $\vec{A}=By\vec{e}_x$ is the vector potential of
a homogeneous magnetic field $\vec{B}= -B \vec{e}_z$, which is perpendicular to the 2DEG, see the inset of
\fig{1}. The Hamiltonian is approximated by finite differences \cite{Gagel1995}
\begin{equation}
  \label{eq:2}
  H_{\mathrm{2DEG}}^{\mathrm{FDA}} =\sum_{\vec{r}_i,\vec{r}_j} t e^{\I \pi \frac{eB}{h} \lr{x_j-x_i}
    \lr{y_j+y_i}} \ket{\vec{r}_i}\bra{\vec{r}_j},
\end{equation}
where the sum is over nearest neighbors in a square lattice of sites at a distance $a=5\un{nm}$, and
$t=\frac{\hbar^2}{2ma^2}\approx 21.8\un{meV}$. This approximation is valid when the magnetic flux through a
unit cell $Ba^2$ is much smaller then a flux quantum $h/e$. We assume that experimentally, the influence of
the temperature is negligible and thus, we set the temperature to $T=0\un{K}$. For simplicity, we also assume
that the spin splitting is not resolved \footnote{This simplification is justified because in GaAs both, the
  effective $g$-factor and the effective mass are $\ll$ 1 and thus, the spin splitting is typically one order
  of magnitude smaller than the Landau splitting.}.

The system size is $800 \un{nm} \times 500 \un{nm}$. Metallic contact regions with a width of $10 \un{nm}$
(corresponding to 3 sites in the finite differences approximation) are attached at the boundaries of the
system separated by a distance $\Dl x= 500\un{nm}$, see the inset of \fig{1}. In order to allow better
comparison of the NEGF calculations with a simplified analytical model, we assume hard-wall boundary
conditions. However, this assumption is not essential for the findings in this paper, see remarks in
Section~\ref{sec:EffDec}. The chemical potential is set to $\mu=0.5t \approx 10.9 \un{meV}$, corresponding to
the carrier density $n_{\mathrm{2D}}\approx 3.3 \cdot 10^{11} \un{cm}^{-2}$. The chemical potential of the
voltage probes $\mu_{P_i}$ is calculated by assuming an infinitesimal bias voltage between $S$ and $D$, and by
using the constraint that voltage measurements are done without a current flow through the voltage probes. The
voltage between $P_1$ and $P_2$ divided by the current between $S$ and $D$ gives the generalized 4-point Hall
resistance $R_{xy}=\frac{\mu_{P_1}-\mu_{P_2}}{eI_{SD}}$ \cite{Houten1989}. Details of the NEGF calculation of
the current and the chemical potential can be found in the following section.

\subsection{Details of the calculations}\label{sec:DetailsCalculations}

Following the NEGF approach, as described in detail by Datta \cite{Datta1997, Datta2005, Datta2012}, the
transmission from the $j$th to the $i$th contact is given by
\begin{equation}
  \label{eq:3}
  T_{ij}= 4\Tr\lr{\Im{\Gm_i}G\Im{\Gm_j}G^+},
\end{equation}
where the Green's function is defined as
\begin{equation}
  \label{eq:4}
  G= \Big[E-H-\textstyle{\sum_{k=1}^{N_c} \Gm_k} \Big]^{-1}.
\end{equation}
The influence of each of the $N_c$ contacts is taken into account by an imaginary self-energy
\begin{equation}
  \label{eq:5}
  \Gm_k= -\I \eta \sum_{\vec{r}_i} \ket{\vec{r}_i}\bra{\vec{r}_i}
\end{equation}
with the broadening constant $\eta= 1.25t\approx 27.25\un{meV}$, representing metallic contact regions. The
sum is over all sites which are coupled to the same contact $k$.

The total current at the $i$th contact is calculated by the Landauer-B\"uttiker formula in its linear response
approximation
\begin{equation}
  \label{eq:6}
  I_i= \frac{2e}{h} \sum_j T_{ij} \lr{\mu_j-\mu_i},
\end{equation}
and the generalized Hall resistance in units of $h/2e^2$ is given by
\begin{equation}
  \label{eq:7}
  R_{xy}= \frac{\sum_j \lr{\mathcal{R}_{P_1j}-\mathcal{R}_{P_2j}}T_{jS}}{T_{DS} +\sum_{ij}T_{Di}\mathcal{R}_{ij}T_{jS}},
\end{equation}
where
\begin{equation}
  \label{eq:8}
  \mathcal{R}^{\,-1}_{ij}=
  \cases{-T_{ij}  & $i \neq j$,\\
    \textstyle{\sum_{k \neq i} T_{ik}}  & $i = j$.\\}
\end{equation}
The sums in \eq{7} are over the contacts with unknown chemical potential, whereas the sum in \eq{8} is over
all contacts including source and drain.

Finite system size effects, such as standing waves between the boundaries of the system, would distort the
focusing spectrum strongly. Therefore, we introduce additional virtual contacts at those boundaries, which are
not essential for the focusing experiment, see the gray sites in the inset of \fig{1}. Mathematically these
virtual contacts can be considered as additional voltage probes with a chemical potential given by the current
conservation constraint. By randomizing the electron phase and momentum \cite{Zilly2009,Stegmann2012}, such
\textit{diffusive walls} mimic an open system and thus, suppress the standing waves. They also greatly reduce
spurious focusing peaks arising from reflections at these boundaries.

The local current of electrons, which originate from the source with energy $\mu$ and which flow from the site
$\vec{r}_j$ to the neighboring site $\vec{r}_i$, is given by \cite{Caroli1971, Cresti2003}
\begin{equation}
  \label{eq:9}
  I_{\vec{r}_i\vec{r}_j}= \frac{2e}{\hbar} \Im{h_{ji}^*A_{ji}^S},
\end{equation}
where the $h_{ij}$ are the matrix elements of the Hamiltonian~\eq{2}. The spectral function for electrons from
the source is defined as
\begin{equation}
  \label{eq:10}
  A^S= -\frac{2}{\pi}G\Im{\Gm_S}G^+.
\end{equation}
The diagonal elements of the spectral function give the local density of states (LDOS), which is accessible to
these electrons.

\section{Properties of the system} \label{sec:Properties} 

\begin{figure}[t]
  \begin{minipage}[c]{0.5\linewidth}
    \includegraphics[scale=0.5]{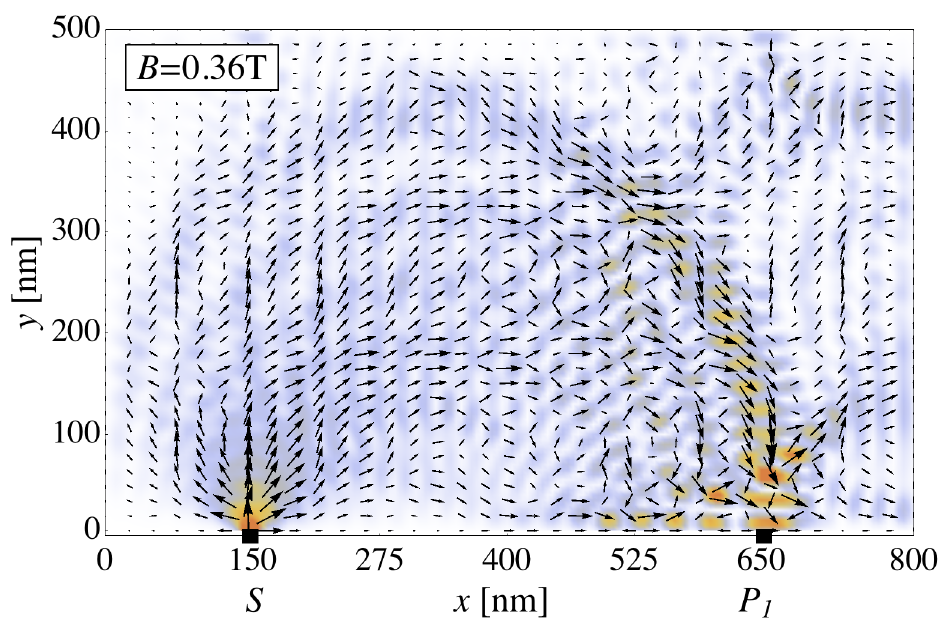}    
  \end{minipage}
  \begin{minipage}[c]{0.5\linewidth}
  \includegraphics[scale=0.49]{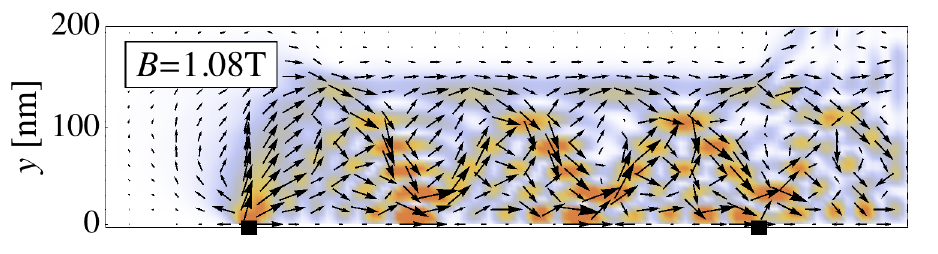}\\
  \includegraphics[scale=0.49]{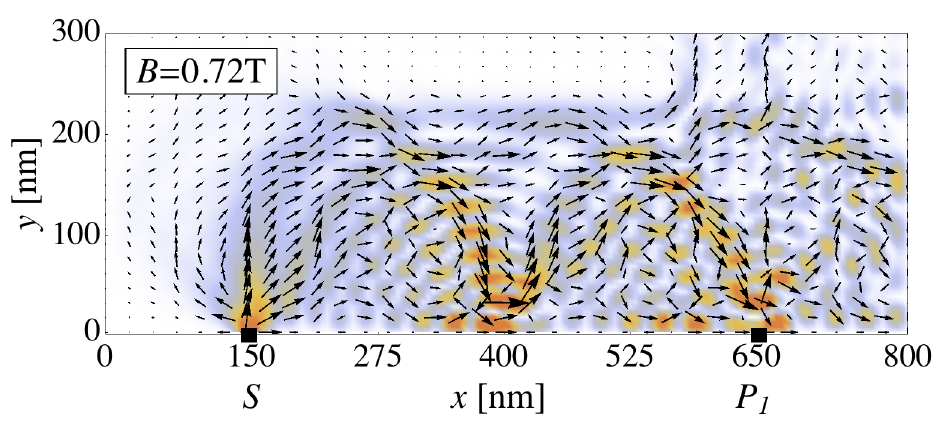}
  \end{minipage}
  \caption{The local current (arrows) and the LDOS (shading) of the electrons originating from $S$ with energy
    $\mu$. The cyclotron orbits can be clearly seen. Also caustics are evident, which are caused by the
    interference of the electrons injected with a broad distribution of angles. Note that at $B= 1.08\un{T}$
    and $B= 0.72\un{T}$ only the relevant part of the system is shown.}
  \label{fig:2}
\end{figure}

The calculated focusing spectrum, i.e. the generalized Hall resistance $R_{xy}$ as a function of the magnetic
field $B$ is shown in \fig{1}. In low magnetic fields ($B < 2\un{T}$), equidistant peaks at
\begin{equation}
  \label{eq:11}
  B_n= \frac{\sqrt{8m\mu}}{e \Dl x} n, \qquad n= 1,2,3 \dots,
\end{equation}
are found (vertical dashed lines). As sketched in the inset, electrons injected by the source $S$ are guided
on cyclotron orbits and end in $P_1$ after $n-1$ reflections at the wall in between, when a multiple of the
cyclotron diameter equals the distance $\Dl x$. These cyclotron orbits can be clearly seen in \fig{2}, which
shows the local current and the LDOS of electrons originating from $S$ with energy $\mu$.

When a diffusive wall is introduced also in between $S$ and $P_1$, the higher focusing peaks are strongly
suppressed, and the extended plateaus of the quantum Hall effect appear, see the dashed curve in \fig{1}. The
Hall resistance is then given by the inverse number of occupied edge channels, which in turn equals the number
of occupied Landau levels. Whenever a Landau level is pushed above the Fermi energy by an increasing magnetic
field, an edge channel vanishes and a step in the resistance can be observed \footnote{In the experiment,
  usually the electron density is constant while the chemical potential is oscillating. However, this would
  only slightly displace the transitions between the Hall plateaus and would not qualitatively change our
  results, see also \cite{Gagel1998}.}.

\begin{figure}[t]
  \begin{minipage}[c]{0.485\linewidth}
    \includegraphics[scale=1]{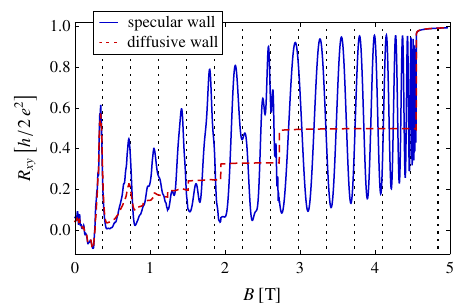}
  \end{minipage}
  \begin{minipage}[c]{0.255\linewidth}
    \includegraphics[scale=0.36]{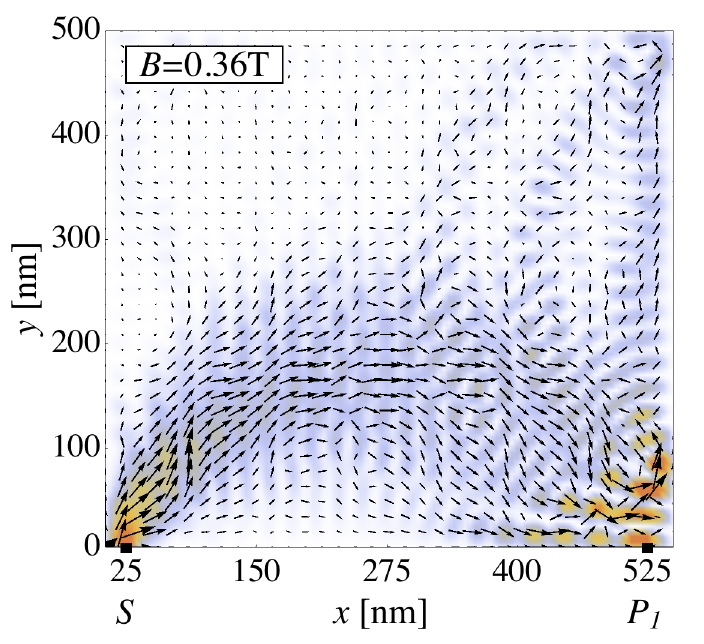}
  \end{minipage}
  \begin{minipage}[c]{0.245\linewidth}
    \includegraphics[scale=0.36]{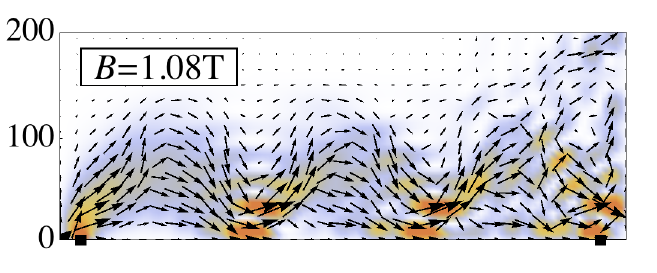}\\
    \includegraphics[scale=0.36]{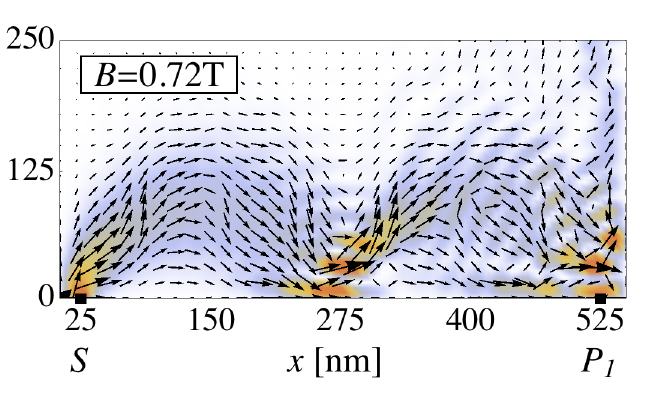}
  \end{minipage}
  \caption{Focusing spectrum and local current, when the distribution of the electron injection angles is
    narrowed by reducing the distance of the injector to the left absorbing wall. Fine structure and caustics
    are suppressed.}
  \label{fig:3}
\end{figure}

As the electrons are injected with a broad distribution of angles, the local current shows caustics
\cite{Houten1989}.  Moreover, interference of the coherent electrons gives rise to a fine structure in the
focusing spectrum and the local current. This can be suppressed, if the distribution of the injection angles
is narrowed by reducing the distance of the injector to the left diffusive wall, see the focusing spectrum and
the local current in \fig{3}.

As expected, \fig{1} shows that the focusing peaks cannot be observed when the direction of the magnetic field
is reversed. The single peak at a low magnetic field is an artefact, which arises when the cyclotron diameter
equals the distance between $S$ and $P_2$.

\subsection{Anomalous properties of the focusing spectrum} \label{sec:AnomalousProperties}

When the strength of the magnetic field is further increased ($B>2\un{T}$), we observe an additional set of
resistance oscillations, which cannot be explained by classical trajectories. The frequency of these
oscillations increases rapidly whenever a Landau level is pushed towards the Fermi energy and a transition
between Hall plateaus appears (compare solid and dashed curves in \fig{1}). Moreover, when only two edge
channels are occupied ($2.7\un{T}<B<4.5\un{T}$), the oscillations become very clear and regular. Finally, the
oscillations vanish completely, when only a single edge channel is occupied ($B>4.5\un{T}$), and the typical
Hall plateau $R_{xy}=1$ can be observed.

This suggests that these oscillations are an interference phenomenon between the occupied edge channels as
explained in the following. We start by solving the stationary Schr\"odinger equation with the Hamiltonian
\eq{1} and an infinite potential wall along the $x$-axis. The Hamiltonian can be rewritten as a harmonic
oscillator, which is shifted by $y_k=\ell^2 k$ with $\ell^2= \frac{\hbar}{eB}$ and thus, the edge channels are
given by \cite{Datta1997}
\begin{equation}
  \label{eq:12}
  \psi_{k,\nu}(x,y)= c_{k,\nu} \, e^{i k x} e^{-\frac{1}{2}\lr{y-y_k}^2/\ell^2} H_{\nu}\Bigl(\frac{y-y_k}{\ell}\Bigr),
\end{equation}
where $c_{k,\nu}$ is a normalization constant. Note that the momentum of the plane wave $e^{ikx}$ parallel to
the infinite wall determines the apex $y_k$ of the parabola. The $H_\nu$ are the Hermite polynoms with index
$\nu$, which is here in general \textit{non-integer} and which is determined numerically by the hard-wall
condition $\psi_{k,\nu}(x,0)=0$.

\begin{figure}[t]
  \centering
  \includegraphics[scale=1.1]{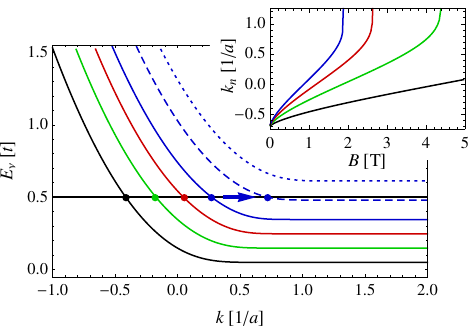}
  \caption{Eigenenergy spectrum \eq{13} of a 2DEG bounded by an infinite potential wall at $y=0$. The solid
    curves give the first four energy bands at $B= 1.3\un{T}$ while the dashed and the dotted curves give the
    fourth energy band at $B=1.8\un{T}$ and $B=2.3\un{T}$, respectively. The dots indicate the $k_n$ at the
    Fermi energy $\mu=0.5 t$. The arrow points out the increase of $k_4$, when the corresponding Landau level
    approaches $\mu$. The inset shows the $k_n$ as a function of $B$.}
  \label{fig:4}
\end{figure}

The \fig{4} shows the resulting eigenenergy spectrum
\begin{equation}
  \label{eq:13}
  E_{\nu}(k)= \hbar \om_c \lr{\nu(k)+1/2},
\end{equation}
with the cyclotron frequency $\om_c=eB/m$. For sufficiently large $k$, the influence of the infinite wall is
negligible and we observe the equidistant Landau levels for integer values of $\nu$. However, the energy bands
are bent upwards, when the apex of the parabola approaches the wall. We numerically calculate the $k_n$, which
agree with a given Fermi energy and a given magnetic field, see the marked intersection points in \fig{4}. The
total number of the $k_n$ gives the number of occupied edge channels.

\begin{figure}[t]
  \centering
  \includegraphics[scale=1.1]{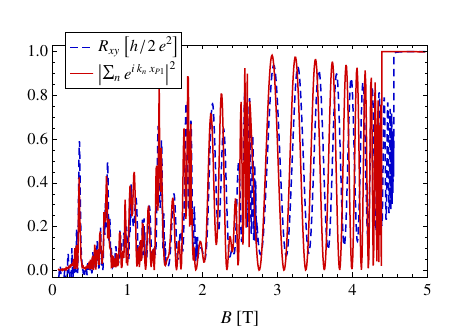}
  \caption{Normalized probability density calculated by a superposition of plane waves with the $k_n$,
    evaluated at the position of the collector $x_{P_1}$ (solid curve). A remarkable agreement with the NEGF
    calculation of \fig{1} (dashed curve) can be observed.}
  \label{fig:5}
\end{figure}

We consider only the plane wave contribution in \eq{12}, which propagates along the infinite wall, and
calculate the superposition of the different $k_n$ with equal weights. The normalized probability density
evaluated at the position of the collector shows remarkable agreement with the NEGF calculation of the
focusing spectrum, see the \fig{5}. Thus, the focusing peaks in low magnetic fields, which correspond to
classical trajectories, can also be explained by the interference of multiple edge
channels \cite{Houten1989,Beenakker1991}. Moreover, this explanation of the focusing spectrum is valid for
every strength of the magnetic field and allows to understand the anomalous peaks.

\begin{figure}[t]
  \centering
  \includegraphics[scale=0.5]{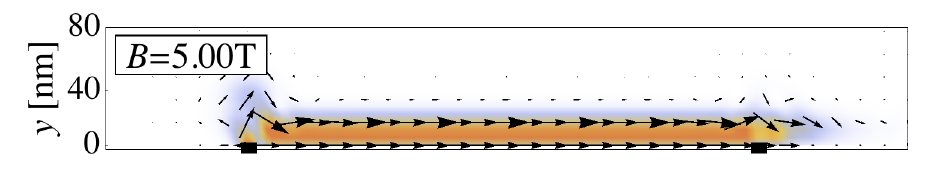}\\
  \includegraphics[scale=0.5]{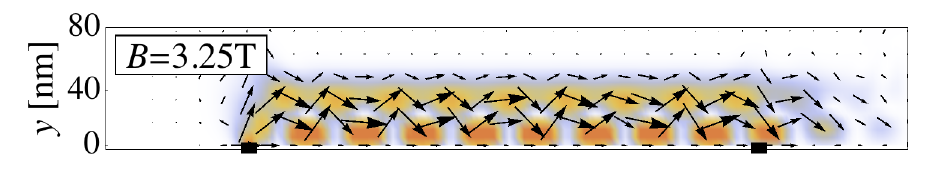}\\
  \includegraphics[scale=0.5]{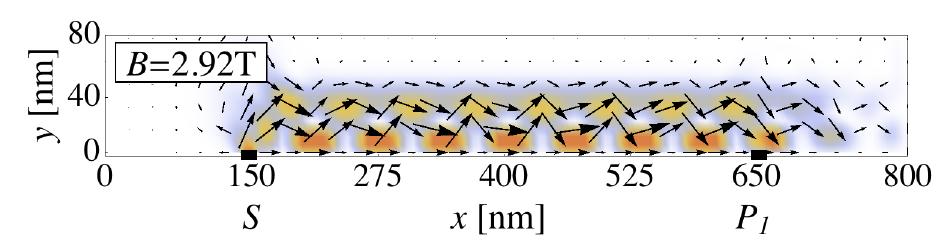}
  \caption{The local current and the LDOS of the electrons originating from $S$ with energy $\mu$. The
    transport through the interfering edge channels in the lower figures resembles to some extent a cyclotron
    motion, while at $B=5 \un{T}$ the current is carried through a single edge channel straight along the
    wall. Only the relevant part of the system is shown.}
  \label{fig:6}
\end{figure}

In intermediate fields, the current is carried by only a few edge channels and the focusing spectrum shows
beatings due to the superposition of plane waves. In particular, when only two edge channels are occupied,
only two plane waves are superimposed and a beating appears, whose frequency is determined by the difference
of $k_1$ and $k_2$. The frequency of the oscillations increases rapidly, whenever the highest occupied Landau
level approaches the Fermi energy, because its intersection point and thus, the corresponding $k_{\mathrm{max}}$
increases strongly, see the divergences in the inset of \fig{4}. Its difference to the other, much smaller
$k_n$ leads to a high frequency beating. Finally, when only a single edge channel is occupied, the beating and
thus, the oscillations in the focusing spectrum vanish. The current then flows along an edge channel parallel
to the wall, see the top of \fig{6}. This figure also illustrates that although the focusing peaks in
intermediate fields cannot be explained by classical trajectories, the local current resembles to some extent
cyclotron motion along the wall.

The clear and distinct oscillations due to the occupation of only two edge channels can also be understood as
a new commensurability between the magnetic flux enclosed within the two edge channels and the flux
quantum. At the maximum of the oscillations the two plane waves interfere constructively, and thus the
difference of their momenta fulfills $\Dl k= 2\pi / \Dl x$. If we relate this momentum difference to the
distance between the edge channels $\Dl y_k= \ell^2 \Dl k$, we obtain
\begin{equation}
  \label{eq:14}
  \Dl x \Dl y_k B = \frac{h}{e}.
\end{equation}
Thus, between two successive focusing peaks, the magnetic flux within the area enclosed by the two edge
channels increases by one flux quantum. In this way, we can relate the focusing spectrum to the distance of
the edge channels and the difference of their momenta. For the experimental observation of such interference
effects with a fixed distance between the edge channels, see e.g. \cite{Deviatov2011} and references therein.

Note that indeed many properties of the system can be understood by a basic quantum calculation. However, this
cannot replace the NEGF approach, which allows to include contacts in a controlled way and to obtain
quantitative results for the Hall resistance. Moreover, the superposition of plane waves with equal weights is
justified only by its good agreement with the NEGF calculation.

\subsection{Effects of decoherence, non-specularity, boundary conditions and contact geometry}
\label{sec:EffDec}

Dynamic scattering like electron-phonon and electron-electron interaction causes decoherence in the system. We
study the effects of weak decoherence on the focusing spectrum by a phenomenological model based on virtual
contacts \cite{Zilly2009, Stegmann2012}. These act as scattering centers, where the electron phase and
momentum are randomized completely, as pointed out by B\"uttiker \cite{Buettiker1986_1, Buettiker1991}. The
virtual contacts are randomly distributed over the system with a relative density $p$. This parameter reflects
the degree of decoherence, ranging from the completely coherent ($p=0$) to the completely incoherent ($p=1$)
case. It is inversely proportional to the phase coherence length. In the finite-difference approximation of
the 2DEG, this is technically done by selecting randomly with probability $p$ bonds, which connect neighboring
sites, and replacing them by virtual contacts. The focusing spectrum is then averaged over multiple
decoherence configurations until convergence is reached.

\begin{figure}[t]
  \begin{minipage}[c]{0.54\linewidth}
    \includegraphics[scale=1.1]{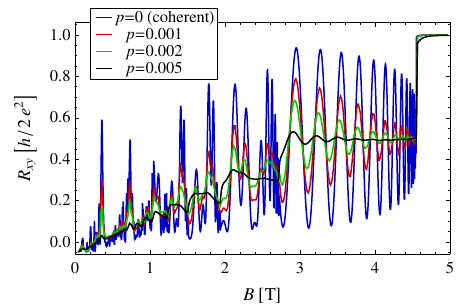}
  \end{minipage}
  \begin{minipage}[c]{0.45\linewidth}
  \includegraphics[scale=0.45]{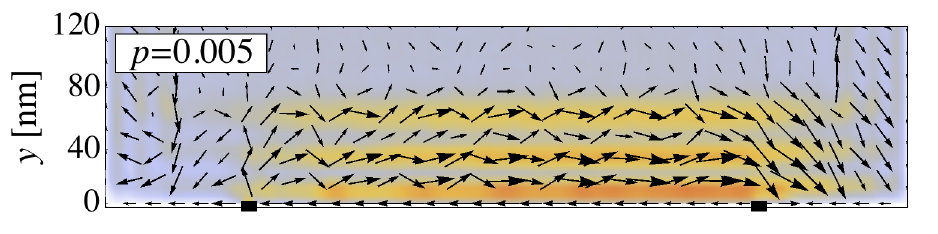}\\
  \includegraphics[scale=0.45]{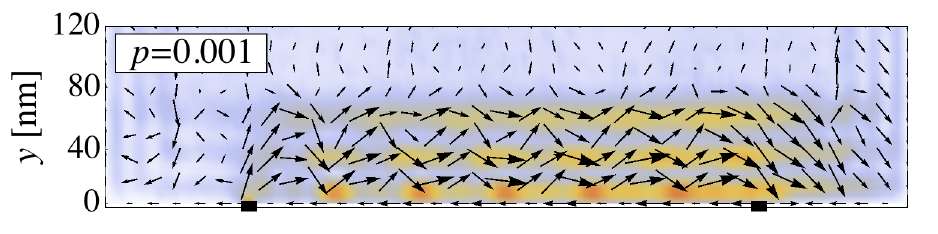}\\
  \includegraphics[scale=0.45]{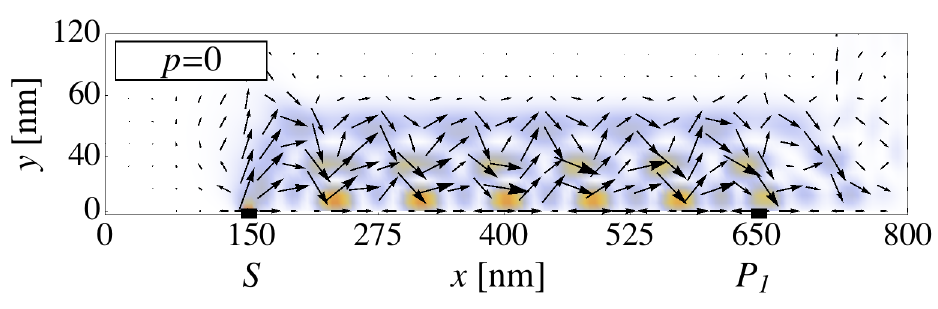}
  \end{minipage}
  \caption{Influence of an increasing degree of decoherence on the focusing spectrum (left) as well as the
    LDOS and local current (right, $B=2.13\un{T}$) for electrons originating from the source. The oscillations
    are gradually suppressed and isolated edge channels remain. Averages are over 75 decoherence
    configurations.}
  \label{fig:7}
\end{figure}

The averaged focusing spectrum in \fig{7} shows that with increasing degree of phase and momentum
randomization all oscillations are suppressed and the surprisingly robust Hall plateaus appear
\cite{Gagel1996b, Gagel1998, Xing2008}. The classical focusing peaks are even stronger suppressed than the
anomalous oscillations, because the latter are located in a much narrower part of the system and thus, are
less influenced by the scattering centers. The LDOS and the local current of electrons originating from the
source show distinct edge states while the cyclotron orbits are vanishing, because the interference between
the edge channels is annihilated by the decoherence. Note that the LDOS is also strongly broadened by the
decoherence. As expected, when the degree of decoherence is further increased ($p > 0.05$) the quantum Hall
plateaus vanish and the classical linear Hall resistance appears.

\begin{figure}[t]
  \centering
  \includegraphics[scale=1.1]{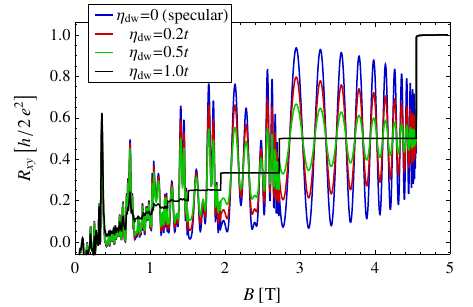}
  \caption{Focusing spectra for an increasingly diffusive boundary. The oscillations in the focusing spectrum
    are suppressed gradually with increasing degree of non-specularity and increasing number of reflections.}
  \label{fig:8}
\end{figure}

Our model also allows us to study the effects of partially specular reflections by introducing between $S$ and
$P_1$ a diffusive wall with the broadening parameter $\eta_{\mathrm{dw}}$. In this way, we can tune the
scattering from specular ($\eta_{\mathrm{dw}}=0 $) to diffusive ($\eta_{\mathrm{dw}}\sim t$). The \fig{8}
shows that the oscillations in the focusing spectrum are suppressed gradually with increasing degree of
non-specularity and increasing number of reflections at the boundary. Moreover, our findings are not dependent
on the chosen boundary conditions (i.e. hard wall). When a parabolic confining edge potential is used,
qualitatively similar results are obtained.

We have also studied the influence of the contact geometry on the focusing spectrum. Using contacts with a
width of $40 \un{nm}$, attached via $120 \un{nm}$ long leads, we found qualitatively the same focusing
spectrum, which clearly shows classical focusing peaks as well as anomalous oscillations.

\subsection{Experimental observability} \label{sec:Exp}

In closing, we discuss requirements to experimentally observe the novel oscillations reported in this
paper. In our calculations we have used parameters ($m=0.07m_e,\: \mu= 10.9 \un{meV}, \: n_{\mathrm{2D}}= 3.3
\cdot 10^{11} \un{cm^{-2}} $) of a high quality 2DEG in a GaAs-AlGaAs heterojunction. We expect that the
omission of the spin splitting will not change the results qualitatively. Figure~\ref{fig:7} shows that the
oscillations can be observed up to a degree of decoherence of $p=0.005$, which corresponds to a phase
coherence length of approximately $1 \un{\mu m}$. Likewise, a distance between $S$ and $P_1$ of $500 \un{nm}$
is easily achievable with today's nanolithography techniques. All this gives us confidence that the predicted
oscillations can indeed be observed experimentally.

\section{Summary} \label{sec:Summary}

We have studied theoretically the coherent electron focusing in a 2DEG with a boundary. In a weak magnetic
field $B$, the Hall resistance $R_{xy}$ shows equidistant peaks, which can be explained by classical
trajectories. In a strong field, an extended plateau $R_{xy}=1$ reflects the quantum Hall effect. Moreover, in
intermediate fields, superimposed on the lower Hall plateaus we find oscillations, which are neither periodic
in $1/B$ (quantum Hall effect) nor periodic in $B$ (classical cyclotron motion).

In general, the focusing spectrum can be understood by the interference of the occupied edge channels. In
intermediate fields only a few edge channels are occupied and their interference causes beatings. The beatings
explain the clear and distinct oscillations in the case of two occupied Landau levels. They constitute a new
commensurability between the flux enclosed within the two edge channels and the flux quantum. The frequency of
the oscillations increases rapidly when a Landau level approaches the Fermi energy because one of the
frequencies contributing to the beating increases strongly. When only a single edge channel is occupied, the
beatings and thus, the oscillations in the focusing spectrum vanish abruptly. Decoherence suppresses the
classical focusing peaks as well as the anomalous oscillations and brings out the quantum Hall plateaus.

\begin{ack}
  This work was supported by Deutsche Forschungsgemeinschaft under Grant No. GRK1240 and No. SPP1386. We are
  grateful to A.~Ganczarczyk, O.~Ujs\'aghy and M.~Zilly for useful discussions and helpful remarks. We thank
  Universit\"atsbibliothek Duisburg-Essen and Deutsche Forschungsgemeinschaft under program ``Open Access
  Publizieren'' for covering the publication charge of this article.
\end{ack}

\section*{References}
\bibliographystyle{unsrt.bst}
\bibliography{./Stegmann-cef}

\end{document}